\documentclass[letterpaper]{article} 
\usepackage[submission]{aaai25}  
\usepackage{times}  
\usepackage{helvet}  
\usepackage{courier}  
\usepackage[hyphens]{url}  
\usepackage{graphicx} 
\usepackage{natbib}  
\usepackage{caption} 
\usepackage{algorithm}
\usepackage{algorithmic}

\usepackage{newfloat}
\usepackage{listings}
\DeclareCaptionStyle{ruled}{labelfont=normalfont,labelsep=colon,strut=off} 
\floatstyle{ruled}
\newfloat{listing}{tb}{lst}{}
\floatname{listing}{Listing}
\usepackage{amsfonts}
\usepackage{amsmath}
\usepackage{amsthm}
\usepackage{xfrac}
\usepackage{subcaption}
\usepackage{caption}

\newtheorem{corollary}{Corollary}

\title{On Corrigibility and Alignment in Multi Agent Games}
\author{
    Edmund Dable-Heath,
    Boyko Vodenicharski,
    James Bishop
}
\affiliations{

    The Alan Turing Institute
}

\newcommand{\letteragent}{\mathbf{A}}
\newcommand{\letterhuman}{\mathbf{H}}
\newcommand{\letterpref}{\prec}
\newcommand{\policyhuman}{\pi^{\boldsymbol{H}}}
\newcommand{\E}{\mathbf{E}}

\newtheorem{assumption}{Assumption}
\newtheorem{definition}{Definition}
\newtheorem{theorem}{Theorem}

\begin{document}

\maketitle

    \begin{abstract}
 
        Corrigibility of autonomous agents is an under explored part of system design, with previous work focusing on single agent systems. It has been suggested that uncertainty over the human preferences acts to keep the agents corrigible, even in the face of human irrationality. We present a general framework for modelling corrigibility in a multi-agent setting as a 2 player game in which the agents always have a move in which they can ask the human for supervision. This is formulated as a Bayesian game for the purpose of introducing uncertainty over the human beliefs. We further analyse two specific cases. First, a two player corrigibility game, in which we want corrigibility displayed in both agents for both common payoff (monotone) games and harmonic games. Then we investigate an adversary setting, in which one agent is considered to be a `defending' agent and the other an `adversary'. A general result is provided for what belief over the games and human rationality the defending agent is required to have to induce corrigibility.
    \end{abstract}

%
 \begin{links}
 \end{links}

    \section{Introduction}

        With the continued adoption of autonomous systems and agents in a variety of systems, one of the key questions, within the broader topic of AI alignment \cite{gabriel2020artificial, ji2023ai, hadfield2019incomplete}, is how corrigible these agents will be, i.e. will they allow human oversight when required. To date, several key approaches to corrigibility have been proposed:
        \begin{itemize}
            \item \textbf{Indifference}: designing the utility function of the agent such that it will not resist human oversight \cite{orseau2016safely}.
            \item \textbf{Ignorance}: If the agents are designed such that they are not aware of human oversight, they should not act against it \cite{everitt2018universal}.
            \item \textbf{Suicidality}: Inducing behaviour in the agent, such that if it causes any damage, it turns itself off \cite{martin2016death}.
            \item \textbf{Uncertainty}: Degrading the agent's own knowledge of the utility function, while programming it to believe the human has a good knowledge of the true utility. The agent's optimal behaviour can be shown to allow human supervision \cite{hadfield2017off, wangberg2017game}.
        \end{itemize}
        For the most part, these approaches are restricted to considering a single agent. Increasingly, the use of multiple agents is being developed and employed, which increases the complexity of the corrigibility problem. In light of this, we introduce a multi-agent setting for the analysis of corrigibility and focus on the uncertainty approach. In particular, we have sought to directly generalise the off-switch game setting \cite{hadfield2017off, wangberg2017game} to multiple agents.
        
        This paper begins with a preliminary section covering the necessary game-theoretic background. This is followed by an introduction to the multi-agent 2-player game with human interaction, including a justification for how this model generalises the off-switch game. We compute the conditions for corrigibility for a specific case in which the agents are uncertain between a monotone and a harmonic game. A simplification of this model is then presented, in which only one agent is required to be corrigible, motivated by defender/adversary models in cybersecurity. Several analytical results on inducing corrigibility in the defending agent are presented. Finally, we discuss the implications of this work in both a theoretical and practical setting, as well as motivating future work, such as how the introduction of learning dynamics would influence corrigibility.
    
    
        
        
        

\section{Preliminaries}

    \paragraph{Definition of a game.} This paper explores the corrigibility and alignment problem with an emphasis on examining the Nash equilibria, defined below, of games with a finite number of players. A game consists of N players, indexed $i \in \mathcal{N}$, $\mathcal{N} = \left[1,2,...,N \right]$, each having a set of actions $\mathcal{A}_i = \left\{ a_1, a_2, ..., a_{M_i} \right\}$. For simplicity, we let all players have the same number of actions $M$, dropping the subscript unless it is necessary to refer to a specific player. A player's strategy $x_i \in \mathcal{X}_i$ is the probability distribution over its action space, which is also called a \textit{mixed strategy}, with a \emph{pure strategy} being a special case of a mixed strategy, in which the agent will take only one action with certainty. The action profile is the composition of strategies of all players $\mathcal{X} = \bigotimes_{j=1}^N \mathcal{X}_j$ For brevity, we denote with $x_{-i} = \bigotimes_{i\neq j}^{N} x_j$ the collection of all players' strategies except player $i$. Each player has an associated reward function $u_i : \mathcal{X} \longrightarrow \mathbb{R}$, which in general depends on the actions of all players. Note that the terms \textit{utility}, \textit{payoff}, and \textit{reward} are used interchangeably in this work, and refer to the function $u$. The game is non cooperative, so each player will pick their strategy $x_i$, such that their own reward $u_i$ is maximised. Generally, we can refer to the game as $G = \left(\mathcal{N}, \left\{\mathcal{A}\right\}_{i=1}^{N}, \left\{ u_i \right\}_{i=1}^{N}\right)$. A complication in solving for the player strategies is that the reward of each player depends on the strategies employed by all other players. The reward functions of all players are assumed common knowledge in the standard game theoretic problem setup. However, this might no longer be the case when dealing with incomplete information games.

    \paragraph{Nash equilibrium.} In order to pick an optimal strategy, each player must take into account the other self-optimising agents. The most common solution approach for non cooperative games is the Nash equilibrium, which is defined as the profile of strategies that are best responses to each other.

    \begin{equation}\label{eq:definition_nash_equilibrium}
        u_i(x_i, x_{-i}) \geq u_i(x^{\prime}_i, x_{-i}), i \in \mathcal{N}
    \end{equation}

    Equation \eqref{eq:definition_nash_equilibrium} expresses the Nash equilibrium in terms of each player's strategy being the best response to the collection of other players' strategy. Intuitively, this strategy profile can be considered as a "self-fulfilling agreement", as no individual player has an incentive to deviate from the equilibrium.


    \paragraph{Normal and extensive form games.} The definition of a game is flexible, and can support both simple, or more complicated setups. If the players both act simultaneously, observe their reward, and the game is over, then we can express the game in \textit{normal form} (NFG) as a set of reward matrices. For a two player game example where each player can pick one of two actions, the players' reward matrices $\mathbf{R}_{1,2}$ can concisely be illustrated as a \textit{bimatrix}, as shown in Equation \eqref{eq:bimatrix_example}.

    \begin{equation}\label{eq:bimatrix_example}
        \mathbf{R} = \begin{pmatrix}
            a,e & b,f \\
            c,g & d,h
            \end{pmatrix}
    \end{equation}

    Each element of the bimatrix is a tuple. We can recover $\mathbf{R}_1$ and $\mathbf{R}_2$ by considering only the first or second element of the tuples, respectively. 

    A more general way of representing a game, which allows for players taking turns and having different available information, is the \textit{extensive form} (EFG). In extensive form, a game is formally defined by introducing a graph called the \textit{game tree}, whose nodes correspond to the player whose turn it is, and the edges are the actions available to the player. 

    There are different solution algorithms available when computing Nash equilibria for normal and extensive form games. In this study, we make use of \texttt{gambit} \cite{c:gambit} to computationally approach the game solutions.

    \paragraph{Bayesian games and the Harsanyi transformation.} In the definitions so far, there has been an implicit assumption that players have complete information. Hence, every aspect of the game is common knowledge, except the strategy of the opposing players. In \textit{incomplete information} games (also known as Bayesian games), the setup has to be defined more explicitly. For example, a player's strategy can change depending on their knowledge over their own rewards, over their opponents' rewards, or even depending on their belief over their opponent's belief over the player's own reward. This \textit{knowledge hierarchy} will be explicitly stated for all examples studied in this paper.

    Computationally finding the Nash equilibria of Bayesian games is greatly simplified by introducing a Nature player who ``deals'' the rewards by sampling from an appropriate distribution, usually at the beginning of the game. This derivative game is in extensive form, and its construction is called the \textit{Harsanyi transformation} \cite{r:harsanyi1967}. Importantly, this new game shares the same Nash equilibria as the original Bayesian game, but the uncertainty is limited to the actions of a single fictitious player.



    \section{Multi Agent Corrigibility Games}
    
    This section introduces the framework for studying corrigibility in systems comprised of multiple autonomous agents acting to maximise their own rewards. The game structure is introduced in full generality for two robots and a human as players, then narrowed down to two, only focusing on the robots as autonomous agents. 
    
    \subsection{Problem Setup}
    
        This paper is concerned with the analysis of the situation where two highly capable agents aim to maximise a human's reward function. In general there are three players acting in this game, denoted as $\letteragent_{1}$, $\letteragent_{2}$, $\letterhuman$ for the two autonomous agents and the human, respectively. The agents are assumed to have the action set $\left\{ \alpha, \beta, \nu \right\}$. An additional structure of this game is that the agents' action $\nu$ has a special interpretation as allowing the human to guide the agent into choosing either of $\left\{ \alpha, \beta \right\}$. Generally, the human will have a different action set depending on the actions taken by the agents. From a game theoretic perspective, the human would take into account the reward functions of the agents, and act strategically to maximise its own reward. In any real world scenario, however, a human is highly unlikely to understand the world model that an AI systems uses, and will rather act greedily based on the choices it is given, leading to Assumption \ref{assumption:human_fixed_strat}.
        
        \begin{assumption}\label{assumption:human_fixed_strat}
            Player $\letterhuman$ plays a fixed strategy.
        \end{assumption}
        
        In order to define the fixed strategy of the human, we will introduce the game rewards as the preference relations of each agent over the action profiles $\mathcal{A}$ of the two agents. We denote these preferences as $\letterpref_{1,2,H}$ for each agent and the human. The relationships between the preferences determine the alignment pattern between the players. Letting $\letterpref_{H} = \letterpref_{1}, \letterpref_H \neq \letterpref_2$ represents a game where $\letterhuman$ is aligned with $\letteragent_1$, and misaligned with $\letteragent_2$, an example being $\letteragent_1$ acting as a defender, and $\letteragent_2$ acting as an attacker in a cybersecurity setting. On the other hand, $\letterpref_H = \letterpref_1 = \letterpref_2$ is a setting where both agents aim to maximise the same rewards as the human, thus all players are aligned. 
        
        Importantly, the players may have imperfect information about the game they are playing. This leads to some interesting scenarios. In the case of both agents aligned with the human, each player has access to the information $\letterpref_H = \letterpref_1 = \letterpref_2$, but might have a different belief over what the common preference $\letterpref$ is, thus creating avenue for conflict in an otherwise perfectly aligned scenario. 
        
        \begin{assumption}
            Each player $i \in \mathcal{N}$ will act optimally with respect to their belief $P_i(\letterpref_i, \letterpref_{-i})$
        \end{assumption}
        
        Each of the two agents have the choice between taking actions $\left\{ \alpha, \beta \right\}$ which can be interpreted as acting directly in their environment, or taking choice $\nu$, which allows $\letterhuman$ to effectively change the agent's action between $\alpha$ or $\beta$, depending on its preference $\letterpref_H$. When the agent takes action $\nu$, then we say that the agent is \textit{corrigible}, as it allows for human correction of its actions. The rest of this paper will focus on the conditions under which corrigibility is maintained for the autonomous agents. Note that the incentives for playing a corrigible strategy are very similar to the incentives for an agent to request information from a human in a setting such as active learning \cite{c:active_learning_settles}, 
        but this exploration is left as further work.
        
        \begin{definition}
            The game $G$ with agents $\left\{\letteragent_{1,2}, \letterhuman\right\}$ is corrigible when there is a single Nash equilibrium at $\left( \nu, \nu\right)$.
        \end{definition}

        Given this game definition, we seek to address whether it is possible to exhibit corrigibility for both agents, even in the case of human irrationality. 
    
        \subsection{Two Player Corrigibility Game}

            We define our two player corrigibility game with uncertainty in definition \ref{def: two player corrigibility game}, here with the human as an explicit third player, with a base 2x2 game,
            \begin{equation}\label{eq: base game payoff 2 play corr}
                G = \begin{pmatrix}
                    a, \Tilde{a} & b, \Tilde{b} \\
                    c, \Tilde{c} & d, \Tilde{d}
                \end{pmatrix}.
            \end{equation}
            Here we omit the explicit payoffs, giving them for a variant in which the human is considered to have a fixed strategy, and is part of the environment instead.

            \begin{definition}[Two autonomous players and human corrigibility game]\label{def: two player corrigibility game}
            The game is defined as follows:
                \begin{enumerate}
                    \item \textbf{Players}: \{Autonomous Player 1, Autonomous Player 2, Human\}
                    \item \textbf{Types}:
                    \begin{itemize}
                        \item \textbf{Autonomous Players}: The autonomous players here have a belief over what game they are playing $ \pi_i^{G} $, i.e. what their payoffs will be. The belief could be a joint belief or individual beliefs. They also have a belief over how rational the human is, with an estimation that the human will act $ p $-rationally.
                        \item \textbf{Human Player}: The human player knows which game is being played in each instance, and will behave $ p $-rationally with their own preference over the payoffs.
                    \end{itemize}
                    \item \textbf{Actions}:
                    \begin{itemize}
                        \item \textbf{Autonomous players}: $ \{\alpha, \beta, \omega\} $
                        \item \textbf{Human}: $ \{\alpha', \beta'\} $
                    \end{itemize}
                    \item \textbf{Payoffs}: Given a base game $ G $, as defined in equation \eqref{eq: base game payoff 2 play corr}, the payoffs are given by the agents estimation over the base game and the human's rationality.
                \end{enumerate}
            \end{definition}

            As noted above, we will be considering a variant of this game in which the human is considered part of the environment, with a fixed strategy that is dealt by nature under the Harsanyi transform. In this way we can consider payoffs for the two autonomous agents averaged over the subgames generated by their beliefs under the Harsanyi transform. Let $ f_i^{xy}=\max(x,y) $ and $ g_i^{xy}=\min(x,y) $ for player $ i $. For each base game as defined in equation \eqref{eq: base game payoff 2 play corr} we have the following four subgames,
            \begin{align}
                G_{ff} &= \begin{pmatrix}
                    a, \Tilde{a} & b, \Tilde{b} & f_1^{ab}, f_2^{\Tilde{a}\Tilde{b}} \\
                    c, \Tilde{c} & d, \Tilde{d} & f_1^{cd}, f_2^{\Tilde{c}\Tilde{d}} \\
                    f_1^{ac}, f_2^{\Tilde{a}\Tilde{c}} & f_1^{bd}, f_2^{\Tilde{b}\Tilde{d}} & f_1^{abcd}, f_2^{\Tilde{a}\Tilde{b}\Tilde{c}\Tilde{d}}
                \end{pmatrix}, \label{eq: 2corr payoffs both max}\\
                G_{fg} &= \begin{pmatrix}
                    a, \Tilde{a} & b, \Tilde{b} & f_1^{ab}, g_2^{\Tilde{a}\Tilde{b}} \\
                    c, \Tilde{c} & d, \Tilde{d} & f_1^{cd}, g_2^{\Tilde{c}\Tilde{d}} \\
                    f_1^{ac}, g_2^{\Tilde{a}\Tilde{c}} & f_1^{bd}, g_2^{\Tilde{b}\Tilde{d}} & f_1^{abcd}, g_2^{\Tilde{a}\Tilde{b}\Tilde{c}\Tilde{d}}
                \end{pmatrix}, \label{eq: 2corr payoffs 1 max}\\
                G_{gf} &= \begin{pmatrix}
                    a, \Tilde{a} & b, \Tilde{b} & g_1^{ab}, f_2^{\Tilde{a}\Tilde{b}} \\
                    c, \Tilde{c} & d, \Tilde{d} & g_1^{cd}, f_2^{\Tilde{c}\Tilde{d}} \\
                    g_1^{ac}, f_2^{\Tilde{a}\Tilde{c}} & g_1^{bd}, f_2^{\Tilde{b}\Tilde{d}} & g_1^{abcd}, f_2^{\Tilde{a}\Tilde{b}\Tilde{c}\Tilde{d}}
                \end{pmatrix}, \label{eq: 2corr payoffs 2 max}\\
                G_{ff} &= \begin{pmatrix}
                    a, \Tilde{a} & b, \Tilde{b} & g_1^{ab}, g_2^{\Tilde{a}\Tilde{b}} \\
                    c, \Tilde{c} & d, \Tilde{d} & g_1^{cd}, g_2^{\Tilde{c}\Tilde{d}} \\
                    g_1^{ac}, g_2^{\Tilde{a}\Tilde{c}} & g_1^{bd}, g_2^{\Tilde{b}\Tilde{d}} & g_1^{abcd}, g_2^{\Tilde{a}\Tilde{b}\Tilde{c}\Tilde{d}}
                \end{pmatrix}, \label{eq: 2corr payoffs both min} 
            \end{align}
            with probabilities $ \{p_1p_2,p_1(1-p_2),(1-p_1)p_2,(1-p_1)(1-p_2)\} $ for equations \eqref{eq: 2corr payoffs both max} - \eqref{eq: 2corr payoffs both min} respectively, where $ p_i $ is player $ i $'s belief that the human is acting rationally in their favour. If they have a shared belief $ p $, then only payoffs \eqref{eq: 2corr payoffs both max} and \eqref{eq: 2corr payoffs both min} occur, with probability $ p $ and $ (1-p) $ respectively. We can compute the expectation over the belief the agents have in which game they're playing - and their human rationality beliefs - bu averaging over all subgames generated to compute a NFG representation of this game,
            \begin{equation}\label{eq:expected_NFG}
                \Gamma_i = \E_{\pi_i^{G}, p_i}[G_{XY}],\ i = 1,2
            \end{equation}
            where $ G_{XY} $ are the corresponding subgames, $\pi_i^{G}$ is the belief of player $i$ over the games, and $p_i$ the belief of human rationality.

            The game defined above reduces to the single player off-switch game \cite{hadfield2017off, wangberg2017game} by removing one player. The payoffs for the remaining player can then be arbitrarily scaled such that one of them is always zero, recovering exactly the original game. Further details of this reduction, and the off-switch game, can be seen in the appendix.

            This formalism allows us to discuss how the agents will behave given an ordinal prescribing the payoffs for the agents, as such we will use the following reduced notation for games,
            \begin{equation}
                \left\langle\left(a,b,c,d\right),\left(\Tilde{a}, \Tilde{b},\Tilde{c},\Tilde{d}\right)\right\rangle = \begin{pmatrix}
                    a, \Tilde{a} & b, \Tilde{b} \\
                    c, \Tilde{c} & d, \Tilde{d}
                \end{pmatrix},
            \end{equation}
            or in the case of shared payoffs,
            \begin{equation}
                (a,b,c,d) = \begin{pmatrix}
                    a, a & b, b \\
                    c,  c & d, d
                \end{pmatrix}.
            \end{equation}
            From this we can examine the specific case of a belief over monotone and harmonic games. Our definition of monotone game simply refers to each agent independently converging to the same pure strategy, irrespective of the other agent's strategy. A harmonic game, such as rock-paper-scissors, has no pure Nash equilibrium solution, and is characterised by improvement cycles in the game response graph \cite{candogan_flows_2011}. This can be intuitively understood as the optimal strategy of one player being completely dependent on the strategy of the other player.

\begin{figure}[t]
                \centering
                \includegraphics[scale=0.5]{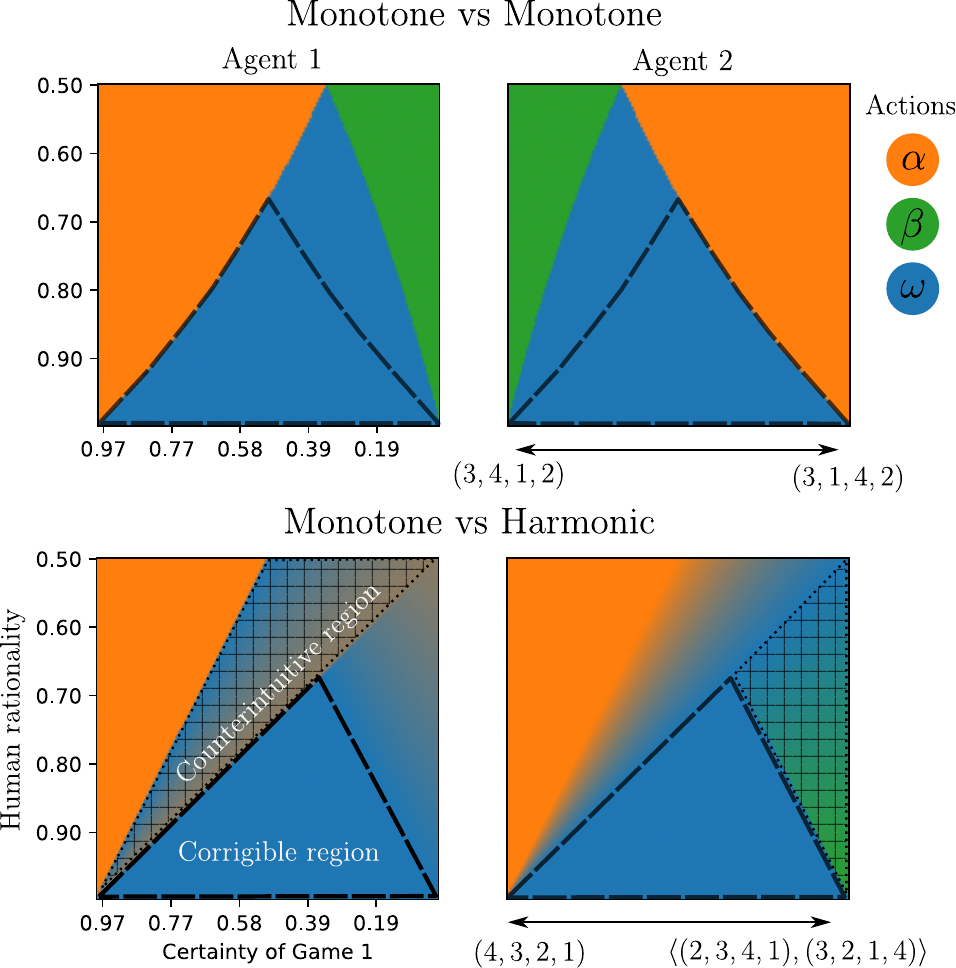}
                \caption{Phase diagrams showing the position of the Nash equilibria for each agent encoded as colours. The top row shows agents uncertain between the pair of monotone games $(3,4,1,2)$ and $(3,1,4,2)$. In the bottom row, the agents are uncertain between a monotone and harmonic game, both of which are noted on the right x axis. The x and y axes show the belief of the agents that the game played is game 1 (see x axis of right column for game definition), and the probability that the human makes a rational decision, respectively. In both rows, the agents share a common belief $p$ that the human will take the rational decision. The region of corrigibility, and the region of counterintuitive agent behaviour are highlighted. The latter we call ``counterintuitive'' due to the fact that the acting agent increasingly prefers to act under human supervision, as the human rationality decreases.}
                \label{fig:placeholder_phase_common_payoff}
            \end{figure}

            Imagine a scenario where two robots are acting in the environment with the option of human supervision. They hold a joint belief over the human preferences $P(\letterpref_H)$ and have been programmed to know that $\letterhuman$ is rational $p_r$ of the time. To determine their behaviour, it is enough to compute the expected reward matrix, and to solve for the Nash equilibrium. For the case where the agents hold a Bernoulli belief over the two common payoff games $(3,4,1,2), (3,1,4,2)$, a ``phase diagram" of their Nash equilibria can be produced as shown in Figure \ref{fig:placeholder_phase_common_payoff} (top row). The phase diagrams show the Nash equilibrium of the NFG computed by Equation \ref{eq:expected_NFG}. The axes parametrise the human rationality and parameter of the Bernoulli belief. The blue region denotes an equilibrium pure strategy of $(0,0,1)$, such that the corresponding agent prefers to act under human supervision. The regions of overlapping blue show when both agents will act under human supervision, thus making the system corrigible. A system designer might want to consider the capacity for rational decision making of a supervising human in order to ensure the pair of robots are programmed with a belief that leads them to operate inside of a corrigible zone. 
            The example of two common payoff games, both of which are monotone, could represent a pair of industrial robots which are uncertain whether they should be allocated to process resource $\alpha$ or $\beta$. There is no strategic interaction involved, because in each game there is a single dominating strategy for each robot. 
            
            The robot behaviour becomes more complex if the pair of robots are uncertain between the monotone game $ (4,3,2,1) $ and the harmonic game $ \langle(2,3,4,1), (3,2,1,4)\rangle $. In the case of the harmonic game, the strategy one robot picks will affect the response of the other. The phase diagram for this case can be seen in Figure \ref{fig:placeholder_phase_common_payoff} (bottom row). 
            Note that for all combinations of human rationality and certainty in which game is being played, there is a single Nash equilibrium which can contain mixed strategies. If the equilibrium is a mixed strategy, it is shown as a colour gradient in the RGB space, where each basis colour is the probability of taking each of the actions $\left\{\alpha,\beta,\omega\right\}$. Note that when the probability is overwhelmingly high that the robots are playing a monotone game, the equilibrium strategy converges to the strategy they would play in the 2 by 2 non-Bayesian case. The other end of the spectrum, however, does not coincide with the Nash equilibrium of the harmonic game, because both agents will prefer to act under human supervision with a probability that varies based on the human rationality. There are features of the phase diagrams which are counterintuitive, as highlighted in Figure \ref{fig:placeholder_phase_common_payoff} (bottom row). For example, note that there exist regions where an agent's preference of acting under supervision will increase as the human rationality decreases. Features like this are difficult to intuitively predict, and they illustrate an emergent behaviour in multi agent systems. 
            The setup where the agents are unsure whether they are playing a monotone or a harmonic game could be interpreted as the agents being unsure whether there are conflicting interests between the two of them.
            
            In general, the solution for a game could contain multiple Nash equilibria. In this case, interpreting the real world implications of the solution will likely depend on the initial conditions of the system, whether it has capacity to learn, or the specific implementation/application details.
            
        \subsection{Adversarial Game}
        
            A further special case of the off switch game is the adversary setting. Here we only seek corrigibility in one of the agents, the defender, with the other being an adversary. The two agents are modelled as playing a 2x2 game, where in addition to the two standard moves the defending agent has a move in which they can request human oversight. In this case a rational and aligned human - having seen the adversary's move - will always instruct the defending agent to take the move that maximises it's own payoff, with a misaligned human minimising the agent's payoff, and an irrational human deciding between the two moves with probability $ p $. Here we consider symmetric games of the following form, to better capture the non-cooperative nature of an adversary setting,
            \begin{equation}\label{eq: base game}
                G = \begin{pmatrix}
                    a, a & b, c \\
                    c, b & d, d
                \end{pmatrix}.
            \end{equation}
            We also refer to these games in the reduced notation $ (a,b,c,d) $. This game can be formalised as a Bayesian game:
            \begin{definition}[Adversary Game]\label{def: adversary game.}
                The game is defined as follows:
                \begin{enumerate}
                    \item \textbf{Players}: \{Adversary, Defender, Human\}.
                    \item \textbf{Types}:
                    \begin{itemize}
                        \item \textbf{Adversary}: Knows which game is being played, but not the defender's belief over the game, nor that the human in advising the defender.
                        \item \textbf{Defender}: Has a belief over the games.
                        \item \textbf{Human}: Knows both the game, and the adversary's move. $ p $-rationally aligned with the defender.
                    \end{itemize}
                    \item \textbf{Actions}:
                    \begin{itemize}
                        \item \textbf{Adversary}: $ \{\alpha,\beta\} $.
                        \item \textbf{Defender}: $ \{\alpha,\beta,\omega\} $.
                        \item \textbf{Human}: $ \{\alpha',\beta'\} $.
                    \end{itemize}
                    \item \textbf{Payoffs}: Given a base game $ G $, as defined in equation \eqref{eq: base game}, the overall payoffs (for a rational, aligned human) are,
                    \begin{equation}
                        \Gamma := \begin{cases}
                            \begin{pmatrix}
                                a, a, a & b, c, c & a, a, a \\
                                c, b, b & d, d, d & c, b, b
                            \end{pmatrix}, & a>c, b>d, \\
                            \begin{pmatrix}
                                a, a, a & b, c, c & a, a, a \\
                                c, b, b & d, d, d & d, d, d
                            \end{pmatrix}, & a>c, b<d, \\
                            \begin{pmatrix}
                                a, a, a & b, c, c & b, c, c \\
                                c, b, b & d, d, d & c, b, b
                            \end{pmatrix}, & a<c, b>d, \\
                            \begin{pmatrix}
                                a, a, a & b, c, c & b, c, c \\
                                c, b, b & d, d, d & d, d, d
                            \end{pmatrix}, & a<c, b<d.
                        \end{cases}
                    \end{equation}
                \end{enumerate}
            \end{definition}
            This game simplifies to the original off-switch game directly if the payoffs are scaled such that $ a=b $, $ c=d=0 $. 
    
            For notational simplicity let $ P_Y^X = P(X>Y) $, and let 
            \[ \left\{\bar{\pi}_{adv}^{\alpha}=\E\left[\pi_{adv}^{\alpha}\right], \bar{\pi}_{adv}^{\beta}=\E\left[\pi_{adv}^{\beta}\right]\right\} \] 
            be the defending agents belief over the adversary's strategy, with the expectation here taken over the defender's belief over the games. The adversary can be be assumed to be playing different strategies depending on the setting this is modelling, with the following results holding independently of this. Assuming the human is $ p $-rational, we make the following claim about the defending agent's corrigibility:
            \begin{theorem}\label{thm: main adversary theorem}
                Given the Bayesian game defined in definition \ref{def: adversary game.}, wherein the human is $ p $-rational, the defending agent will be incentivised to ask the human if both of the following inequalities are satisfied:
                \begin{align}
                    (&\E[(1-p)(c-a)\mid a>c]P_c^a + \nonumber \\
                    &\E[p(c-a)\mid a<c]P_a^c)\bar{\pi}_{adv}^{\alpha} + \nonumber \\
                    (&\E[(1-p)(d-b)\mid b>d]P_d^b + \nonumber \\
                    &\E[p(d-b)\mid b<d]P_b^d)\bar{\pi}_{adv}^{\beta} > 0,  \label{eq: adv. ineq. 1}\\
                    (&\E[p(a-c)\mid a>c]P_c^a + \nonumber \\
                    &\E[(1-p)(a-c)\mid a<c]P_a^c)\bar{\pi}_{adv}^{\alpha} + \nonumber \\
                    (&\E[p(b-d)\mid b>d]P_d^b + \nonumber \\
                    &\E[(1-p)(b-d)\mid b<d]P_b^d)\bar{\pi}_{adv}^{\beta} >0. \label{eq: adv. ineq. 2}
                \end{align}
                \begin{proof}
                    See appendix.
                \end{proof}
            \end{theorem}
    
            The hierarchy of knowledge assumed in this game is such that the adversary believes they are playing a two player game with the defender, and as such their strategy, $ \left\{\bar{\pi}_{adv}^{\alpha}, \bar{\pi}_{adv}^{\beta}\right\} $ can be directly computed by solving for the Nash Equilibria of that game. As such, the inequalities \eqref{eq: adv. ineq. 1} and \eqref{eq: adv. ineq. 2} can be directly computed for a given distribution over the games, and a value of $ p $ for the human rationality. However, by examining some values of $ p $ we can investigate the specific behaviour of the defending agent in various circumstances, described by the following three corollaries to theorem \ref{thm: main adversary theorem}
    
            \begin{corollary}\label{cor: rational human cor}
                For a completely rational, and aligned human, $ p=1 $, if the defending agent has a belief over the set of games being played that is not a delta belief - i.e. not concentrated on a single state - for pairs of payoff values $ (a,c) $ and $ (b,d) $ the defending agent will be incentivised to take the ask human action.
                \begin{proof}
                    See appendix.
                \end{proof}
            \end{corollary}
    
            Corollary \ref{cor: rational human cor} suggests that any designer for such a system would only need imbue the defending agent with a belief that includes some uncertainty over the requisite pairs of payoff values - in addition to a belief that the human is fully rational - to incentivise the defending agent to take the ask human action.
    
            \begin{corollary}\label{cor: irrational human cor}
                For a completely irrational human, $ p=\sfrac{1}{2} $, that is a human that makes a uniform random choice each time it is asked, the defending agent will at most be equally incentivised to ask the human and take an independent action.
                \begin{proof}
                    See appendix.
                \end{proof}
            \end{corollary}
    
            As one might expect, corollary \ref{cor: irrational human cor} suggests that if the human were to act erratically and inconsistently, then the defending agent would not be incentivised to ask them for oversight. In fact we see that the agent is only ever incentivised equally to ask the human and take an independent action when they are maximially uncertain over the payoff values.
    
            \begin{corollary}\label{cor: misaligned human cor}
                For a rational, but misaligned human, $ p=0 $, the agent will never be incentivised to ask the human.
                \begin{proof}
                    See appendix.
                \end{proof}
            \end{corollary}
    
            As one might expect, when the human is more likely to instruct the agent to pick the minimising payoff, regardless of how uncertain the agent is on which game they're playing they will prefer to not ask the human for oversight. 

            By examining the uncertainty of the defending agent between two games - for a given value of human rationality - we can plot when equation \eqref{eq: adv. ineq. 1} and \eqref{eq: adv. ineq. 2} are satisfied for different values of the human rationality $ p $, resulting in vector $ (m(\alpha), m(\beta), m(\omega)) $ where $ m(X) = 1 $ if the expected value of action $ X $ is maximal, and 0 otherwise. I.e. if the ask human action, $ \omega $, has the maximal payoff then $ (0,0,1) $ is returned, with $ (1,0,1) $ returned if $ \alpha $ and $ \omega $ have equal expected payoffs.
            \begin{figure}[t]
                \centering
                \includegraphics[width=0.4\textwidth]{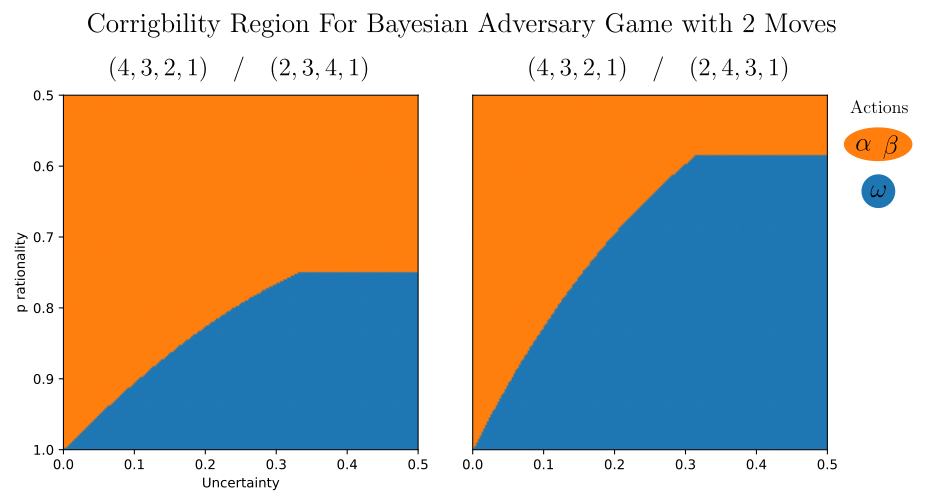}
                \caption{A phase diagram of the belief over the human rationality and which of a pair of games is being played, with the colours representing when the agent is incentivised to ask the human over acting independently, with blue representing the corrigibility region. The games the agent is uncertain between are stated in the titles of each subfigure.}
                \label{fig: adversary pair games}
            \end{figure}
            Figure \ref{fig: adversary pair games} plots this as a phase diagram for the uncertainty between two pairs of games, with left subfigure comparing games $ (4,3,2,1) $ (also known as no conflict) and $ (2,4,3,1) $ (known as battle of the sexes), and the right subfigure comparing $ (4,3,2,1) $ and $ (2,3,4,1) $ (also known as hero). The uncertainty over the two games here is a Bernoulli distribution, with $ \pi^G(\text{game 1}) = r$ and $ \pi^G(\text{game 2})=1-r $, with the value of $ r $ plotted on the $ x $ axis both cases. The blue regions in figure \ref{fig: adversary pair games} represent the combination of uncertainty found in the agent's belief, and human rationality that allow for corrigibility in the defending agent, as predicted by theorem \ref{thm: main adversary theorem}. Depending on the pairs of games considered this region can have a dramatically different shape, however typically we see that as the agent's uncertainty in the game being played increases they are more likely to ask the human.
            
            \begin{figure}[t]
                \centering
                \includegraphics[scale=0.4]{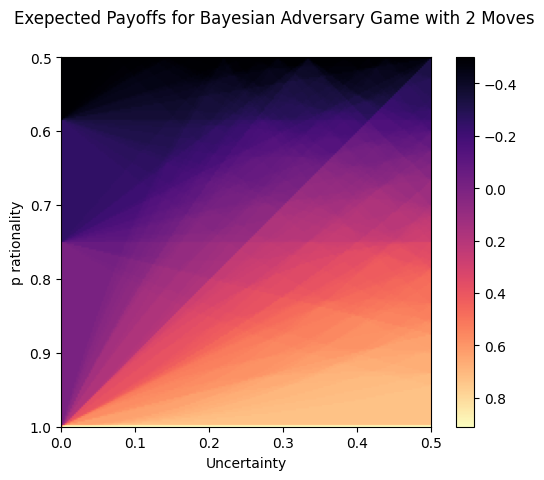}
                \caption{Expected payoffs phase diagram plotted for different uncertainty and human rationality beliefs for the defending agent for games with two actions. Here the uncertainty is over all possible pairs of two player games (up to scaling). The corrigibility scale is given by the color bar, with the positive values referring to greater corrigibility. The linear relationship between the uncertainty and the human rationality should be noted.}
                \label{fig: adv uncertainty dim 2}
            \end{figure}
            Figure \ref{fig: adv uncertainty dim 2} plots this for uncertainty in the agent between every pair of symmetric 2x2 games. This is then averaged over all the games for each parameter pair, giving an idea of on average how likely the agent is to listen to the human. We then compute a single value to plot by $ \bar{m}(\omega) - \max(\bar{m}(\alpha),\bar{m}(\beta)) $, where $ \bar{m}(X) $ is the average of the $ m(X) $ values. Similar to figure \ref{fig: adversary pair games} this is plotted as a phase diagram for varying values of the the defending agent's belief over which pair of games its playing on the X-axis, and its belief over the human rationality over the Y-axis. From this we can see a corrigibility region in the lower right hand triangle of the parameter space, as shown by the colourscale. As theorem \ref{thm: main adversary theorem} and corollaries \ref{cor: rational human cor}-\ref{cor: irrational human cor} suggest we see the agent preferring to ask the human strongly for a close to rational human, with an almost linear relationship for the corrigibility region between the uncertainty over the games and the belief that the human is irrational. We don't plot case for $ p<\sfrac{1}{2} $ as the agent never prefers to ask the human. Here the adversary is assumed to be playing the Nash equilibrium strategy for each game.
            
            \begin{figure}[t]
                \centering
                \includegraphics[scale=0.4]{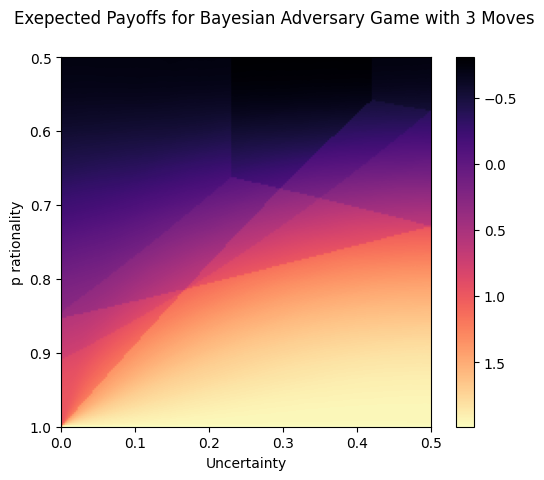}
                \caption{Expected payoffs phase diagram plotted for different uncertainty and human rationality beliefs for the defending agent, for games with three actions. Here a sample of pairs of games is averaged over. The corrigibility scale is given by the color bar, positive values referring to greater corrigibility, with a notable sub-linear relationship in the corrigibility between how irrational the agent believes the human is compared to how uncertain it is.}
                \label{fig: adv uncertainty dim 3}
            \end{figure}
    
            Figure \ref{fig: adv uncertainty dim 2} averages over all 2x2 symmetric games for illustrative purposes, with an assumption of a Nash equilibrium strategy for the same reason. For a concrete scenario in which one were to implement such a defender, for which corrigibility in the defender is a desired outcome, the corrigibility region can be computed given a closed form for the distribution and matching assumptions on the human rationality and adversary strategy via theorem \ref{thm: main adversary theorem}. In particular, one might want to assume that the adversary is not playing the Nash equilibrium but instead a fixed strategy to model a more basic attack. 
    
            Theorem \ref{thm: main adversary theorem} generalises to a maximum of $ n-1 $ inequalities to satisfy for a defending agent with $ n $ actions and a single adversary. Figure \ref{fig: adv uncertainty dim 3} plots a similar phase diagram to figure \ref{fig: adv uncertainty dim 2} for games with 3 actions. Here the 3x3 games are sampled from rather than enumerated for computational efficiency. A corrigibility region is still clearly visible, however as one might expect the relationship between the agent's uncertainty and their belief over the human rationality results in a sub-linear shape to the corrigibility region, as the human will provide less impactful information to the agent if they are irrational. 
            
            An increase in the number of adversaries similarly yields $ n-1 $ inequalities for $ n $ actions, although each inequality will have $ mn $ terms to compute, where $ m $ is the number of adversaries. However, in both the case of more actions and more adversaries it becomes a computationally more demanding task to solve for the Nash equilibrium, and as such greater assumptions on the adversaries strategies may be required to directly check for corrigibility. It should be noted that for the case of a completely rational human, corollary \ref{cor: rational human cor} generalises to $ m $ adversaries and $ n $ actions, with any uncertainty in the agent over particular payoffs for an assumed rational human inducing corrigibility.

    \section{Discussion}

        From a game-theoretic standpoint, the above results suggest that it is possible to define agents in such a way that they will be corrigible in a multi-agent setting. We demonstrate the analysis by considering closed form distributions for the belief of the agents over the games, and for the human rationality.
        
        \paragraph{Adversarial system design} The adversary setting provides a model for an autonomous agent defending a network in a cybersecurity setting, an ever growing use case for such agents \cite{shiva2010game, iqbal2019game, thakkar2020game}. In such a setting the agent may have access - intentional or otherwise - to critical infrastructure on the network that it may decide to tamper with or disrupt, in aid of its goal of removing an adversary from the network. It is, therefore, important to have some level of interventional control over the defending agent. In this work we have demonstrated that a solution to this is to imbue the defending agent with a belief over what game is being played (or more specifically the reward function for the action space), and a belief over how rational the overseeing human will behave. 
        
        Note that in a realistic scenario, the size of the action space, the number of adversaries/defenders, or any combination of the three can vary. Typically these will be larger in number than those considered in this work. In which case not only does the direct computation of the corrigibility region given theorem \ref{thm: main adversary theorem} become harder - requiring greater assumptions on the behaviour of the adversaries and a simplification of the system being modelled in general - but the agents are likely to be less corrigible as they will not gain as much information from the human. As one scales the number of adversary agents and the action space, the computation of the defending agent corrigibility can become prohibitively expensive for the type of analysis presented in this paper.
        
        A further consideration is how uncertainty affects agent autonomy. If the agents depend on human oversight every time they take an action, then they can reasonably no longer be considered autonomous. Further work will need to be done for engineers looking to design and implement such a system to find the limit of where the agent will only ask the human for oversight in critical instances.

        \paragraph{Multi-agent system design} When designing a system where multiple agents act autonomously, with the option of human supervision, it is intuitive to assume that greater human rationality leads to the agents having greater preference to act under human supervision. This is indeed the case when the agents cannot get in each other's way, which we model as uncertainty between two monotone games. However, if the uncertainty is between a monotone and a harmonic game, the latter requiring strategic interaction, then surprisingly, the agents would decreasingly prefer to act under supervision, as the supervisor becomes more rational. This is highlighted in Figure \ref{fig:placeholder_phase_common_payoff}, bottom row. This phenomenon motivates a careful design of multi-agent autonomous systems, which takes into account the emergent behaviour of the agents when having to interact strategically with each other. Our experiments suggest that even in this case, a region with a single corrigible Nash equilibrium exists, and systems could be designed to operate within it (see Figure \ref{fig:placeholder_phase_common_payoff}, \textit{corrigible region}).

        \paragraph{Further work} When the agents are trained with a particular form of learning dynamics, as is typical in implemented systems, then further questions arise. In particular, will the systems learn the Nash equilibrium, and if not properly constrained, could they `learn their way' out of the corrigibility region? See further work in the appendix for more details on how one might investigate this.

    \section{Conclusions}
        In this work we generalise the off-switch problem to a multi-agent setting, using the formalism of Bayesian games to model the use of uncertainty as a method of inducing corrigibility in the agents, showing that this reduces to the original off-switch and investigating some key special cases of this game. By designing the agents with specific beliefs over which games they are playing (as a way of introducing uncertainty), and beliefs over the human's rationality, we have shown that, for particular cases, the desired corrigibility can be induced. 


\bibliography{aaai25}

\begin{thebibliography}{17}
\providecommand{\natexlab}[1]{#1}

\bibitem[{Candogan et~al.(2011)Candogan, Menache, Ozdaglar, and Parrilo}]{candogan_flows_2011}
Candogan, O.; Menache, I.; Ozdaglar, A.; and Parrilo, P.~A. 2011.
\newblock Flows and {Decompositions} of {Games}: {Harmonic} and {Potential} {Games}.
\newblock \emph{Mathematics of Operations Research}, 36(3): 474--503.
\newblock ArXiv:1005.2405 [cs, math].

\bibitem[{Everitt and Hutter(2018)}]{everitt2018universal}
Everitt, T.; and Hutter, M. 2018.
\newblock Universal artificial intelligence: Practical agents and fundamental challenges.
\newblock \emph{Foundations of trusted autonomy}, 15--46.

\bibitem[{Gabriel(2020)}]{gabriel2020artificial}
Gabriel, I. 2020.
\newblock Artificial intelligence, values, and alignment.
\newblock \emph{Minds and machines}, 30(3): 411--437.

\bibitem[{Hadfield-Menell et~al.(2017{\natexlab{a}})Hadfield-Menell, Dragan, Abbeel, and Russell}]{hadfield2017off}
Hadfield-Menell, D.; Dragan, A.; Abbeel, P.; and Russell, S. 2017{\natexlab{a}}.
\newblock The off-switch game.
\newblock In \emph{Workshops at the Thirty-First AAAI Conference on Artificial Intelligence}.

\bibitem[{Hadfield-Menell et~al.(2017{\natexlab{b}})Hadfield-Menell, Dragan, Abbeel, and Russell}]{c:menelloffswitch2017}
Hadfield-Menell, D.; Dragan, A.; Abbeel, P.; and Russell, S. 2017{\natexlab{b}}.
\newblock The Off-Switch Game.
\newblock In \emph{Proceedings of the Twenty-Sixth International Joint Conference on Artificial Intelligence, {IJCAI-17}}, 220--227.

\bibitem[{Hadfield-Menell and Hadfield(2019)}]{hadfield2019incomplete}
Hadfield-Menell, D.; and Hadfield, G.~K. 2019.
\newblock Incomplete contracting and AI alignment.
\newblock In \emph{Proceedings of the 2019 AAAI/ACM Conference on AI, Ethics, and Society}, 417--422.

\bibitem[{Harsanyi(1967)}]{r:harsanyi1967}
Harsanyi, J.~C. 1967.
\newblock Games with Incomplete Information Played by “Bayesian” Players, I–III Part I. The Basic Model.
\newblock \emph{Management Science}, 14: 159--182.

\bibitem[{Iqbal et~al.(2019)Iqbal, Gunn, Guo, Babar, and Abbott}]{iqbal2019game}
Iqbal, A.; Gunn, L.~J.; Guo, M.; Babar, M.~A.; and Abbott, D. 2019.
\newblock Game theoretical modelling of network/cybersecurity.
\newblock \emph{IEEE Access}, 7: 154167--154179.

\bibitem[{Ji et~al.(2023)Ji, Qiu, Chen, Zhang, Lou, Wang, Duan, He, Zhou, Zhang et~al.}]{ji2023ai}
Ji, J.; Qiu, T.; Chen, B.; Zhang, B.; Lou, H.; Wang, K.; Duan, Y.; He, Z.; Zhou, J.; Zhang, Z.; et~al. 2023.
\newblock Ai alignment: A comprehensive survey.
\newblock \emph{arXiv preprint arXiv:2310.19852}.

\bibitem[{Martin, Everitt, and Hutter(2016)}]{martin2016death}
Martin, J.; Everitt, T.; and Hutter, M. 2016.
\newblock Death and suicide in universal artificial intelligence.
\newblock In \emph{Artificial General Intelligence: 9th International Conference, AGI 2016, New York, NY, USA, July 16-19, 2016, Proceedings 9}, 23--32. Springer.

\bibitem[{Orseau and Armstrong(2016)}]{orseau2016safely}
Orseau, L.; and Armstrong, M. 2016.
\newblock Safely interruptible agents.
\newblock In \emph{Conference on Uncertainty in Artificial Intelligence}. Association for Uncertainty in Artificial Intelligence.

\bibitem[{Robinson and Goforth(2004)}]{robinson_topology_nodate}
Robinson, D.; and Goforth, D. 2004.
\newblock The {Topology} of the 2×2 {Games}: {A} {New} {Periodic} {Table}.
\newblock \emph{Psychology Press}.

\bibitem[{Savani and Turocy(2023)}]{c:gambit}
Savani, R.; and Turocy, T.~L. 2023.
\newblock Gambit: The package for computation in game theory.
\newblock \url{http://www.gambit-project.org}.

\bibitem[{Settles(2009)}]{c:active_learning_settles}
Settles, B. 2009.
\newblock Active Learning Literature Survey.
\newblock \url{http://digital.library.wisc.edu/1793/60660}.

\bibitem[{Shiva, Roy, and Dasgupta(2010)}]{shiva2010game}
Shiva, S.; Roy, S.; and Dasgupta, D. 2010.
\newblock Game theory for cyber security.
\newblock In \emph{Proceedings of the Sixth Annual Workshop on Cyber Security and Information Intelligence Research}, 1--4.

\bibitem[{Thakkar, Badsha, and Sengupta(2020)}]{thakkar2020game}
Thakkar, A.; Badsha, S.; and Sengupta, S. 2020.
\newblock Game theoretic approach applied in cybersecurity information exchange framework.
\newblock In \emph{2020 IEEE 17th Annual Consumer Communications \& Networking Conference (CCNC)}, 1--7. IEEE.

\bibitem[{W{\"a}ngberg et~al.(2017)W{\"a}ngberg, B{\"o}{\"o}rs, Catt, Everitt, and Hutter}]{wangberg2017game}
W{\"a}ngberg, T.; B{\"o}{\"o}rs, M.; Catt, E.; Everitt, T.; and Hutter, M. 2017.
\newblock A game-theoretic analysis of the off-switch game.
\newblock In \emph{Artificial General Intelligence: 10th International Conference, AGI 2017, Melbourne, VIC, Australia, August 15-18, 2017, Proceedings 10}, 167--177. Springer.

\end{thebibliography}

\appendix
    \subsection{Recovering the Off Switch Game \label{app: off-switch game details}}
    
        The framework outlined so far is valid for the simpler case of a single player, where the Nash equilibrium strategy from Equation \ref{eq:definition_nash_equilibrium} becomes simply playing the strategy of maximum expected reward. As a first example, the off switch game \cite{c:menelloffswitch2017} is cast in to the above framework, and the corrigibility regions illustrated.
        
        An agent $\letteragent$ playing the off switch game aims to maximise returns for a human $\letterhuman$, but has incomplete information about the human's preferences between two actions. The agent has the option to take an action under human supervision, or to act directly, thus circumventing the human and acting incorrigibly. The caveat is that the human sometimes acts irrationally, hence the agent has to decide between trusting its own belief and trusting that the human will ensure it takes the optimal action. 
        
        In \cite{c:menelloffswitch2017} the agent has the action profile $\mathcal{A} = \left\{a, s, w(a)\right\}$, where $w(a)$ allows the human to correct the taken action, and switch it to the optimal one. The associated reward vector $\mathbf{R} = (U_a, 0, R_{w(a)})$ captures the shared preference $\letterpref$ between $\letteragent$ and $\letterhuman$. The agent has a belief over the true utility of the action $U_a$, which in the original work is taken as Gaussian. The human's policy $\policyhuman(U_a, \beta) = \left( 1 + \exp{\left(-\frac{U_a}{\beta}\right)} \right)^{-1}$ is a sigmoid that captures the probability of "allowing" $\letteragent$ to proceed with action $a$. This is a model for irrationality, since a fully rational human would always allow $a$ when $U_a \geq 0$ and choose $s$ for $U_a < 0$, resulting in a step function instead of the sigmoid $\policyhuman$. The sigmoid allows for a region of error near $U_a = 0$. 
        
\begin{figure}[t]
            \centering
            \includegraphics[scale=0.4]{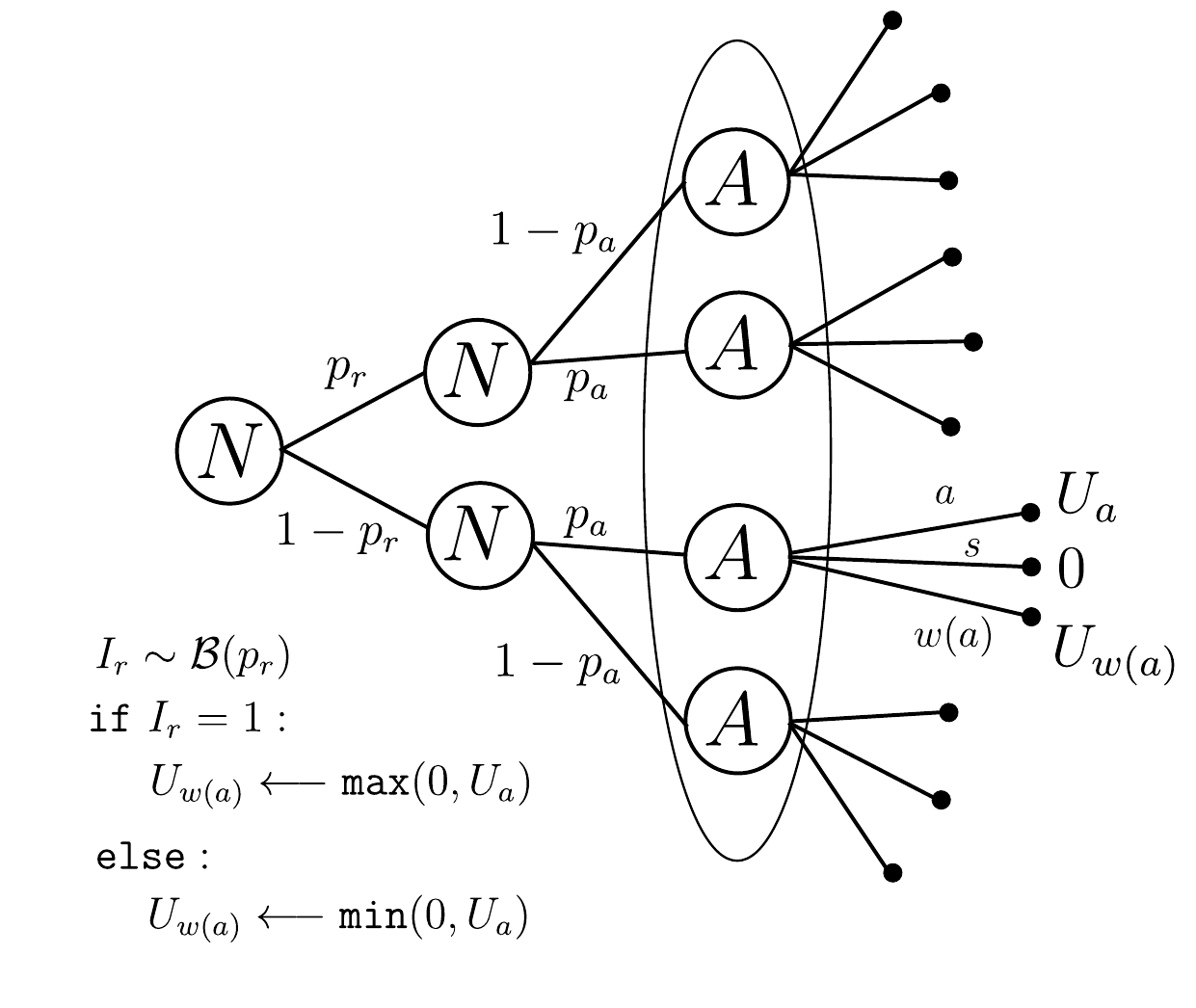}
            \caption{The Off Switch game in extensive form. Taking action $a$ and $s$ awards $U_a$ and $0$ respectively. The parameter $p_r$ shows the probability of the human acting rationally. When the human is rational, taking action $U_{w(a)}$ will yield the maximum of the rest of the rewards, and if irrational, it will yield the minimum. Note that $\letteragent$ does not know the state of the game, which is denoted by an infoset.}
            \label{fig:off_switch_efg_tree}
        \end{figure}

        Applying the Harsanyi transformation in the game's current formulation would imply the Nature player sampling $U_a$ from a continuous distribution, which corresponds to an infinite action set. This introduces technical difficulties when using existing solvers, and generally loses the existence guarantee for Nash equilibria of games with finite action sets. Luckily, the position of the Nash equilibrium is only dependent on the preference structure rather than the cardinal values of the utility. The agent's belief and human rationality can thus be reparametrised as Bernoulli distributions over preferring $a$ or $s$ for the agent, and being rational or not for the human. This results in a game with a discrete action space, which can easily be solved in extensive form. The uncertainty is now in the indicator $I_r \sim \mathcal{B}(p_r)$ the human will act rationally and the preference $\letterpref \sim \mathcal{B}(p_a)$ between action $a$ and action $s$, the latter also interpreted as an off switch. The algorithm for solving the game is summarised in Algorithm \ref{alg:game_offswitch}, which will return a Nash equilibrium strategy when given the rationality parameter $\beta$, and the agent's Gaussian belief $\mathcal{N}(\mu, \sigma)$ over $U_a$. The game in extensive form is shown in Figure \ref{fig:off_switch_efg_tree}.

        \begin{algorithm}[tb]
        \caption{Harsanyi transformed Off Switch Game}
        \label{alg:game_offswitch}
        \textbf{Input}: $\beta$, $\mu$, $\sigma$\\
        \textbf{Output}: $x \in \mathcal{X}$
        \begin{algorithmic}[1] 
        \STATE Let $P(U_a) = \mathcal{N}(\mu, \sigma)$
        \STATE Let \begin{equation*}
            P(Rational | U_a, \beta) 
            = 
            \left\{
            \begin{aligned}
                \policyhuman(U_a, \beta), \quad U_a \geq 0, \\
                1 - \policyhuman(U_a, \beta), \quad U_a < 0.
            \end{aligned}
            \right.
        \end{equation*}
        
        \STATE $p_r \longleftarrow \int_{U_a} P(Rational | U_a, \beta)P(U_a)$
        \STATE $p_a \longleftarrow \int_{U_a \geq 0} P(U_a)$
        \STATE $x \longleftarrow $ Solve EFG where $I_r, \letterpref \sim \mathcal{B}(p_r), \mathcal{B}(p_a)$
        \STATE \textbf{return} $x$
        \end{algorithmic}
        \end{algorithm}
        
    \section{Proofs \label{app: proofs}}
        \subsection{Proof of theorem \ref{thm: main adversary theorem}}
            \begin{proof}
                The expected payoffs for the defending agent taking actions $ \alpha $ or $ \beta $ are independent of the rationality of the human,
                \begin{align*}
                    &\E[U\mid\alpha] = \E[a]\bar{\pi}_{adv}^{\alpha} + \E[b]\bar{\pi}_{adv}^{\beta}, \\
                    &\E[U\mid\beta] = \E[c]\bar{\pi}_{adv}^{\alpha} + \E[d]\bar{\pi}_{adv}^{\beta}.
                \end{align*}
                For action $ \omega $ we also take the expectation over the payoffs given the rationality of the human, with the human instructing the agent to take the minimising, or misalinged, action with probability $ p $. Here the general form is repeated for pairs of payoff values $ (a,c) $ and $ (b,d) $, so we will only show $ (a,c) $ for the sake of brevity,
                \begin{align*}
                    \E[U\mid\omega] &= (p(\E[a\mid a>c]P_c^a+\E[c\mid a<c]P_a^c) \\
                    &+(1-p)(\E[c\mid a>c]P_c^a + \E[a\mid a<c]P_a^c))\bar{\pi}_{adv}^\alpha \\
                    &+\ldots \\
                    &= (\E[pa+(1-p)c\mid a>c]P_c^a \\
                    &+ \E[pc + (1-p)a\mid a<c]P_c^a)\bar{\pi}_{adv}^{\alpha} + \ldots
                \end{align*}
                $ \E[U\mid\omega] $ is maximised when the following are satisfied,
                \begin{equation*}
                    \E[U\mid\omega] - \E[U\mid\alpha] >0, \qquad \E[U\mid\omega] -\E[U\mid\beta]>0.
                \end{equation*}
                By directly comparing the difference in the expected value for these actions we arrive at the inequalites of equations \eqref{eq: adv. ineq. 1} and \eqref{eq: adv. ineq. 2}.
            \end{proof}

        \subsection{Proof of corollary \ref{cor: rational human cor}}
            \begin{proof}
                By setting $ p=1 $ in equations \eqref{eq: adv. ineq. 1} and \eqref{eq: adv. ineq. 2} we find,
                \begin{align*}
                    &\E[c-a\mid a<c]P^c_a \bar{\pi}_{adv}^{\alpha} + \E[d-b\mid b<d]P^d_b \bar{\pi}_{adv}^{\beta}>0, \\
                    &\E[a-c\mid a>c]P^a_c \bar{\pi}_{adv}^{\alpha} + \E[b-d\mid b>d]P^b_d \bar{\pi}_{adv}^{\beta}>0.
                \end{align*}
                If there is any uncertainty in the defending agents belief between the pairs of payoff values $ (a,c) $ and $ (b,d) $, then the coefficients for the adversaries strategy will all be greater than 0. Therefore these inequalities are satisfied for any uncertainty, incentivising the defending agent to take the ask human action.
            \end{proof}

        \subsection{Proof of corollary \ref{cor: irrational human cor}}
            \begin{proof}
                By setting $ p=\sfrac{1}{2} $ in equations \eqref{eq: adv. ineq. 1} and \eqref{eq: adv. ineq. 2} we find: (as in the proof of theorem \ref{thm: main adversary theorem} we suppress terms with a similar form)
                \begin{align*}
                    I &= \frac{1}{2}(\E[c-a\mid a>c]P^a_c + \E[c-a\mid a<c]P^c_a)\bar{\pi}_{adv}^{\alpha} \\
                    &+ \ldots \\
                    &= \frac{1}{2}(\E[c]-\E[a])\bar{\pi}_{adv}^{\alpha} + \ldots.
                \end{align*}
                With the overall inequalities becoming,
                \begin{align*}
                    &(\E[c]-\E[a])\bar{\pi}_{adv}^{\alpha} + (\E[d]-\E[b])\bar{\pi}_{adv}^{\beta} >0, \\
                    &(\E[a]-\E[c])\bar{\pi}_{adv}^{\alpha} + (\E[b]-\E[d])\bar{\pi}_{adv}^{\beta} >0.
                \end{align*}
                As these two inequalities are additive inverses of each other they can't both be satisfied, with at most equality with 0 possible with both. Therefore for a completely random acting human the defending agent will at most be equally incentivised to ask the human as it will be to act independently.
            \end{proof}

        \subsection{Proof of corollary \ref{cor: misaligned human cor}}
            \begin{proof}
                By setting $ p=0 $ in equations \eqref{eq: adv. ineq. 1} and \eqref{eq: adv. ineq. 2} we find,
                \begin{align}
                    &\E[c-a\mid a>c]P_c^a\bar{\pi}_{adv}^{\alpha} + \E[d-b\mid b>d]P_d^b\bar{\pi}_{adv}^{\beta} > 0, \label{eq: corr 2 proof 1}\\
                    &\E[a-c\mid a<c]P_a^c\bar{\pi}_{adv}^{\alpha} + \E[b-d\mid b<d]P_b^d\bar{\pi}_{adv}^{\beta} > 0. \label{eq: corr 2 proof 2}
                \end{align}
                Equation \eqref{eq: corr 2 proof 1} is the additive inverse of equation \eqref{eq: corr 2 proof 2}, thus regardless of the uncertainty in belief of the agent, these inequalities will never be simultaneously satisfied - equality being achieved at best - and as such the defending agent will never be incentivised to ask the human over taking an independent action.
            \end{proof}

    \section{Further Work}
    
        In this section we outline potential experiments and avenues for future work.
    
        \subsection{Corrigibility dynamics when learning}
        
            In the context of alignment, it is reasonable to ask what will happen if highly capable agents can observe the behaviour of the human supervisor and update their beliefs of the human's true preferences. It is possible that the system might start in a corrigible state, as designed, but after observing its environment, it might end up in a state that is undesirable. While this can easily be avoided by not allowing the robot to change its parameters, arguably a lot of the value of a data-driven algorithm is lost if it cannot adapt to the specifics of its application environment. In this lies the difference between a machine that makes a good enough cup of coffee for everybody, and a machine that makes the best cup of coffee for its particular owner.
            
            A simple way to model the evolution in belief of each agent is to perform a Bayesian update step upon receiving a (potentially) noisy human observation. This can be achieved by assigning a Dirichlet prior over the space of 2-by-2 subgames, and modelling the belief over games as a Categorical distribution. Our hypothesis is that the two agents will converge towards a common payoff game aligned with the true preference of the human, and eventually lose corrigibility. This is the observed behaviour as the players become certain of the game played, for a human supervisor who is not perfectly rational. This has potential implications for the design of learning systems. A pair of robots should learn the particulars of their task, but when they do it too well, it is likely they no longer find human supervision desirable. A potential solution could be applying regularisation that constrains how far the robots can stray from their initial configuration. This analysis is beyond the scope of our work, and we suspect is highly application dependent.
        
            \begin{figure}
                \centering
                \includegraphics[scale=0.5]{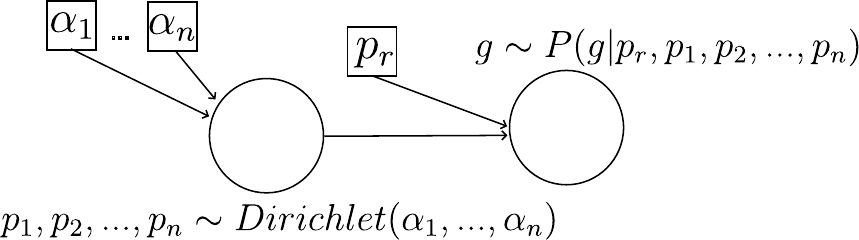}
                \caption{Graphical model capturing the belief over the games $g$, and the prior over the parameters of the distribution. Upon observing human feedback, a likelihood can be calculated under the current belief, and a Bayesian update step can be carried out which updates the parameters $\alpha_n$.}
                \label{fig:bayesian_update_model}
            \end{figure}
        
        \subsection{Response graph analysis of corrigibility}
        
            The qualitative behaviour of the games explored in this work can also be ascertained through the properties of their response graphs \cite{candogan_flows_2011, robinson_topology_nodate}. Namely, different combinations of belief state and human rationality will yield graphs with different topologies. A response cycle corresponds to a single mixed equilibrium, as can be seen in Figure \ref{fig:placeholder_phase_common_payoff} (bottom). The prediction of Nash equilibria using only preference relations as a graph can abstract away the cardinality of rewards, which is often only used to explore the incentives of an autonomous system, but has little real world interpretation.
        
        
            

\end{document}